\documentstyle[preprint,prb,tighten,aps,epsf]{revtex}
\begin{document}
\draft
\title{Spin effects in ferromagnetic single-electron transistors}
\author{J. Barna\'s}
\address{Department of Physics, Adam Mickiewicz University,\\
 ul. Umultowska 85, 61-614 Pozna\'n, Poland}
\date{Received \hspace{5mm} June 1996}
\date{\today}
%\thanks{e-mail:}
\author {J. Martinek, G. Micha{\l}ek, B. R. Bu{\l}ka}
\address{Institute of Molecular Physics, Polish Academy of Sciences,\\
 ul. Smoluchowskiego 17, 60-179 Pozna\'n, Poland}
\author {A. Fert}
\address{Unite Mixte de Physique CNRS/Thomson, 91-404 Orsay, France}

%\thanks{e-mail:}
\date{Received \hspace{5mm} }
\date{\today}
\maketitle

\begin{abstract}

Electron tunneling in ferromagnetic single-electron transistors is considered
theoretically in the sequential tunneling regime. A new formalism
is developed, which operates in a two-dimensional space of states, instead of one-dimensional
space used in the spinless case.
It is shown that spin fluctuations can be
significantly larger than the charge fluctuations.
The influence of discrete energy spectrum of a small central electrode
on tunneling current, charge and spin accumulation, charge and spin
fluctuations, and on tunnel magnetoresistance is analyzed in details.
Two different scales are found in the bias dependence of the basic transport
characteristics; the shorter one originates from the discrete energy spectrum
and the longer one from discrete charging of the central electrode.
The features due to
discrete spectrum and discrete charging disappear at high temperatures.

\end{abstract}


\section{INTRODUCTION}

Electron tunneling in ferromagnetic junctions is of current
interest due to expected applications in magnetic storage
technology and in other spin-electronics devices. Most of
experimental and theoretical works published up to now deal with
tunnel magnetoresistance (TMR) in simple planar junction, i.e.,
with variation of the junction resistance when magnetic
configuration of the junction is changed. Tunneling in more
complex junctions, particularly in mesoscopic ones, where charging
effects are important, was studied only very recently. A special
kind of such junctions are double junctions with a small central
electrode (called alternatively island in the following).
Tunneling in such junctions, known also as Single Electron
Transistors (SET's), was extensively studied in the past decade,
but only in the nonmagnetic limit.$^{1}$ It was shown that when
the electrical capacitance $C$ of the central electrode is small
enough, the charging energy $E_c=e^2/2C$ can be larger than the
thermal energy $k_BT$ and  discrete charging of the central
electrode can lead to Coulomb blockade of electric current below a
certain threshold voltage and to characteristic 'Coulomb
staircase' at higher voltages. However, the interplay of
ferromagnetism and discrete charging was studied only very
recently.$^{2-7}$ It has been shown that discrete charging can
lead to oscillations in TMR.$^{4}$ In Ref.[4] the intrinsic spin
relaxation time on the central electrode was assumed to be
sufficiently short (of the order of the time
between successive tunneling events or shorter) to neglect spin accumulation.
Apart from this, quantization of energy levels of the central
electrode was neglected and the considerations were restricted to
the limit where orthodox tunneling theory is applicable, i.e., to
the case where the barrier resistances are larger than the quantum
resistance $R_Q,\: R_Q=e^2/\hbar$. In that limit higher order
processes (cotunneling) can be generally neglected, except in the
Coulomb blockade regime, where they can play an important role and
can significantly enhance TMR.$^6$ When the intrinsic spin
relaxation time on the central electrode is sufficiently long
(much longer than the time between successive tunneling events),
spin accumulation on this electrode has to be taken into account
and can lead to new phenomena.$^5$ First, the spin accumulation
can enhance TMR. It can also generate TMR when the central
electrode is nonmagnetic. Second, it can give rise to a negative
differential resistance. Third, it can reverse sign of the tunnel
magnetoresistance.

Quantized nature of energy spectrum of a small central electrode
and fluctuations in the spin accumulation were ignored in the works on magnetic
SET's done up to now.
These restrictions are relaxed in the present paper, where both
energy level quantization and spin fluctuations are taken explicitly into
account. Some preliminary results have been published elsewhere.$^8$
Accordingly, we consider a double junction in which all
three electrodes can be ferromagnetic. In a general case relative
orientation of  magnetic moments of the three electrodes
can be arbitrary. When the three electrodes have different coercive fields,
shape anisotropy and/or some of them are exchange biased, then the magnetic
configuration can be easily controlled by a small external magnetic field.
However, we
restrict our considerations to the case where the magnetization of
one of the external electrodes and of the island are parallel
to one another and parallel (parallel configuration) or antiparallel
(antiparallel configuration) to the magnetization of the second external
electrode.
General geometry of the junction considered in this paper
is shown schematically in Fig.1.

In Section 2 we describe the formalism used for calculating
electric current, junction resistance and other characteristics
of the system.
Numerical results are presented and discussed in Section 3. Summary and
final conclusions are in Section 4.

\section{DESCRIPTION OF THE METHOD}

The formalism described in this section is a generalization of the
formalism developed for spinless SET's.$^{9,10,11}$ We consider a
double junction in which the external electrodes are
ferromagnetic, while the central one is either magnetic or
nonmagnetic. The junction is shown schematically in Fig.1, where
spin dependent discrete energy levels of the central (magnetic)
electrode are also indicated. When the central electrode is
nonmagnetic, the energy levels are spin degenerate. Generally, we
assume that the left  and central electrodes have parallel
magnetizations, while the magnetic moment of the right electrode
can be changed from antiparallel to parallel alignment (e.g. by
applying an external magnetic field), as indicated in Fig.1. A
bias voltage $V$ is applied in such a way that the right (left)
electrode is the source (drain) electrode for electrons. A gate
voltage $V_G$ is applied capacitively to the central electrode
(not shown in Fig.1). Apart from this, we assume that electron
spin is conserved during tunneling through the barriers and the
spin dependent resistances of the left ($R_{l\sigma}$) and right
($R_{r\sigma}$) junctions are larger than the quantum resistance
$R_Q$.

Let $E_{i_\sigma}$ denote the single-electron
energy levels of the central electrode at $V=0$.
The index $i_\uparrow$ runs over all energy levels
for spin $\sigma =\uparrow$, while the
index $i_\downarrow$ runs over all energy levels
for spin $\sigma =\downarrow$.
The discrete energy levels $E_{i_\sigma}$
include contributions from all magnetic and nonmagnetic
interactions within the central electrode, like electron correlations
responsible for ferromagnetism, magnetic anisotropy, etc (the Zeemann
term is neglected as the magnetic field assumed to control magnetic
configuration is assumed to be small). Generally, the discrete levels
depend on the number
of electrons in the central electrode
and on their distribution. In our
description, however, we simplify the problem and assume that the discrete levels
are independent of the electron distribution, so the energy spectrum moves
'rigidly' up or down when a bias voltage is applied
and/or when the central electrode becomes charged with a certain number of excess
electrons.$^{10}$ This approximation
is reasonable when the total number of electrons on
the central electrode is significantly  larger than the number
of excess electrons and larger than the number of spins accumulated
on the centeral electrode.

When a bias voltage $V$ is applied, then a stationary electric current
flowing through the junction is then given by
\begin{eqnarray}\label{3}
I=e \sum_{\sigma} \sum_{i_\sigma} \sum_{\{n\}} \Gamma^l_{i_\sigma} P(\{n\})
\{ \delta [n_{i_\sigma},1] [1-f(E_{i_\sigma}+E^{l-}_{N^\star}-E_F)] \nonumber \\
-\delta [n_{i_\sigma},0] f(E_{i_\sigma}+E^{l+}_{N^\star}-E_F) \} \;
=-e \sum_{\sigma} \sum_{i_\sigma} \sum_{\{n\}} \Gamma^r_{i_\sigma} P(\{n\}) \nonumber \\
\times\{ \delta [n_{i_\sigma},1] [1-f(E_{i_\sigma}+E^{r-}_{N^\star}-E_F)]
-\delta [n_{i_\sigma},0] f(E_{i_\sigma}+E^{r+}_{N^\star}-E_F) \} \;,
\end{eqnarray}
where $\{n\}$ denotes a particular
distribution of the occupation numbers,
$\{n\}\equiv\{n_{\uparrow};n_{\downarrow}\}
\equiv\{n_{1_\uparrow},...,n_{i_\uparrow},.
..;n_{1_\downarrow},...,n_{i_\downarrow},...\}$,
of the energy levels $E_{i_\sigma}$,
with $n_{i_\sigma}=1$
($n_{i_\sigma}=0$) when the energy level $i_\sigma$ is
occupied (empty).
$P(\{n\})$ is the stationary probability of
the configuration $\{n\}$ while $\delta [n,n^\prime]$ is defined as
$\delta [n,n^\prime]=1$ for $n=n^\prime$ and
$\delta [n,n^\prime]=0$ for $n\ne n^\prime$.
Apart from this, $e$ denotes the electron charge ($e>0$),
$E^{l\pm}_{N^\star}$ and $E^{r\pm}_{N^\star}$ are defined as
$E^{l\pm}_{N^\star}=eV^{l}_{N^\star}\pm E_c$ and $E^{r\pm}_{N^\star}
=-eV^{r}_{N^\star}\pm E_c$, where
$E_c=e^2/(2C)$ is the charging energy, $N^\star$ is the number of excess
electrons on the central electrode, and $V^l_{N^\star}$ ($V^r_{N^\star}$) is the
electrostatic potential drop on the left (right) junction,
\begin{equation}
V^l_{N^\star}=\frac{ C_r+C_G}{C}V+\frac{N^\star e}{C}-\frac{C_G}{C}V_G\;,
\end{equation}
\begin{equation}
V^r_{N^\star}=\frac{ C_l}{C}V-\frac{N^\star e}{C}+\frac{C_G}{C}V_G.
\end{equation}
Here,  $C_l$ and $C_r$ denote capacitance of left and right
junctions, respectively, $C_G$ is
the gate capacitance, and $C$ is the total capacitance of the central electrode,
$C=C_l+C_r+C_G$.
When writing Eq.(1) we also assumed Fermi-Dirac distribution function,
$f(E-E_F)$,
of the charge carriers in the external electrodes, with $E_F$ denoting
the Fermi level (as in Fig.1). Finally,
$\Gamma^l_{i_\sigma}$ ($\Gamma^r_{i_\sigma}$) in Eq.(1) is the
tunneling rate of electrons from the left
(right) electrode to the level $E_{i_\sigma}$ of the island,
\begin{equation}
\Gamma^{l(r)}_{i_\sigma}={2\pi\over\hbar}\vert M_{i_\sigma}^{l(r)}\vert^2D^{l(r)}_\sigma \; ,
\end{equation}
where $M_{i_\sigma}^{l(r)}$ is an average matrix element
for transitions from the left (right) electrode to the level
$i_\sigma$ and $D^{l(r)}_\sigma$ is the spin dependent density
of electron states in the left (right) electrode.
We assumed above that the charging energy $E_c$ is independent of
the number of electrons on the central electrode and on their
distribution. This is usual approximation within the 'orthodox'
description of single electron tunneling. This approximation is valid for
thermalized distribution of electrons in the central electrode.
When the electrons
on the central electrode are not in thermal equilibrium,
then the charging energy depends on a particular distribution of the electrons,
as shown recently,
both experimentally$^{12}$ and theoretically.$^{13}$
Taking into account the assumption of partial thermalization
of electrons at the central electrode, as will be described later,
we assume $E_c$ to be constant.

The number of electrons with spin $\sigma$ on the central
electrode is equal $N_{\sigma}=\sum_{i_\sigma} n_{i_\sigma}$ and
the total number of electrons is $N=N_{\uparrow}+N_{\downarrow}$.
It is convenient for future analysis to introduce also the number
of excess electrons of a given spin orientation $\sigma$ as,
$N_\sigma^\star = N_\sigma -N_{0\sigma}$, where $N_{0\sigma}$ is
the number of electrons with spin $\sigma$ in equilibrium (at
$V=0$). Note that $N^\star =N_\uparrow^\star +
N_\downarrow^\star$. Magnetic moment of the island is then
determined by the number $M=N_{\uparrow}-N_{\downarrow}$, while
the excess magnetic moment by the number $M^\star =M-M_0$, where
$M_0$ is the equilibrium value of the number $M$ at $V=0$,
$M_0=N_{0\uparrow}-N_{0\downarrow}$.

The probability  $P(\{n\})$  can be determined from a stationary
solution of the following master equation:
\begin{eqnarray}\label{4}
\frac{\partial P(\{n\})}{\partial t}=0=
-\sum_{\sigma} \sum_{i_\sigma}P(\{n\})A(i_\sigma|\{n\})\nonumber\\
+\sum_{i_\uparrow} P(\{n_{1_\uparrow},...,n_{(i-1)_\uparrow},
n_{i_\uparrow}=1 ,n_{(i+1)_\uparrow} ,...;n_{\downarrow}\})
B(i_\uparrow|\{n\})\; \nonumber\\
+\sum_{i_\downarrow}P(\{n_{\uparrow};n_{1_\downarrow},...,n_{(i-1)_\downarrow},n_{i_\downarrow}=1 ,n_{(i+1)_\downarrow} ,...\})B(i_\downarrow|\{n\}) \; \nonumber\\
+ \sum_{i_\uparrow}P(\{n_{1_\uparrow},...,n_{(i-1)_\uparrow},n_{i_\uparrow}=0 ,n_{(i+1)_\uparrow} ,...;n_{\downarrow}\})C(i_\uparrow|\{n\}) \; \nonumber\\
+\sum_{i_\downarrow}P(\{n_{\uparrow};n_{1_\downarrow},...,n_{(i-1)_\downarrow},n_{i_\downarrow}=0 ,n_{(i+1)_\downarrow} ,...\})C(i_\downarrow|\{n\}) \; \nonumber\\
-\sum_{\sigma} \sum_{\sigma^\prime}\sum_{i_\sigma}
   \sum_{j_{\sigma^\prime}}P(\{n\})H(i_\sigma ,j_{\sigma^\prime}|\{n\}) \; \nonumber\\
+\sum_{i_\uparrow , j_\uparrow}P(\{n_{1_\uparrow},...,n_{(i-1)_\uparrow},n_{i_\uparrow}=1 ,n_{(i+1)_\uparrow} ,...,n_{(j-1)_\uparrow},n_{j_\uparrow}=0 ,n_{(j+1)_\uparrow} ,...;n_{\downarrow}\}) D(i_\uparrow,j_\uparrow|\{n\}) \; \nonumber\\+
\sum_{i_\downarrow , j_\downarrow}P(\{n_{\uparrow};n_{1_\downarrow},...,n_{(i-1)_\downarrow},n_{i_\downarrow}=1 ,n_{(i+1)_\downarrow} ,...,n_{(j-1)_\downarrow},n_{j_\downarrow}=0 ,n_{(j+1)_\downarrow} ,...\})D(i_\downarrow,j_\downarrow|\{n\})
\; \nonumber\\+
 \sum_{i_\uparrow, j_\downarrow}\Bigl[P(\{n_{1_\uparrow},...,n_{(i-1)\uparrow},n_{i\uparrow}=1 ,n_{(i+1)_\uparrow} ,...;n_{1_\downarrow},...,n_{(j-1)_\downarrow},n_{j_\downarrow}=0 ,n_{(j+1)_\downarrow} ,...\})S(i_\uparrow,j_\downarrow|\{n\})\nonumber\\
    +P(\{n_{1_\uparrow},...,n_{(i-1)_\uparrow},n_{i_\uparrow}=0 ,n_{(i+1)_\uparrow} ,...;n_{1_\downarrow},...,n_{(j-1)_\downarrow},n_{j_\downarrow}=1 ,n_{(j+1)_\downarrow} ,...\})
S(j_\downarrow,i_\uparrow|\{n\})\Bigr]\;,
\end{eqnarray}
where
\begin{eqnarray}\label{5}
A(i_\sigma|\{n\})=\delta [n_{i_\sigma},0] \left\{\Gamma^l_{i_\sigma} f(E_{i_\sigma}+E^{l+}_{N^\star} -E_F)+\Gamma^r_{i_\sigma} f(E_{i_\sigma}+E^{r+}_{N^\star}-E_F) \right\}\nonumber\\
+\delta [n_{i_\sigma},1] \left\{\Gamma^l_{i_\sigma} [1-f(E_{i_\sigma}+E^{l-}_{N^\star}-E_F)]+\Gamma^r_{i_\sigma}[1-f(E_{i_\sigma}+E^{r-}_{N^\star}-E_F)] \right\}\;, \\
B(i_\sigma|\{n\})=\delta [n_{i_\sigma},0]
\times\Biggl\{\Gamma^l_{i_\sigma} [1-f(E_{i_\sigma}+E^{l-}_{N^\star +1}-E_F)] \nonumber \\
+\Gamma^r_{i_\sigma}[1-f(E_{i_\sigma}+E^{r-}_{N^\star +1}-E_F)] \Biggr\} \;\\
C(i_\sigma|\{n\})= \delta [n_{i_\sigma},1] \left\{\Gamma^l_{i_\sigma} f(E_{i_\sigma}+E^{l+}_{N^\star -1}-E_F)+\Gamma^r_{i_\sigma} f(E_{i_\sigma}+E^{r+}_{N^\star -1}-E_F) \right\}\;,\\
H(i_\sigma ,j_{\sigma^\prime }|\{n\})= \delta [\sigma ,\sigma^\prime ]
\delta [n_{i_\sigma},0]\; \delta [n_{j_\sigma},1]\; d_{j_\sigma,i_\sigma}+ \delta [\sigma ,-\sigma^\prime ]\delta [n_{i_\sigma},0]\; \delta [n_{j_{-\sigma}},1] \;w_{j_{-\sigma },i_\sigma }\;,\\
D(i_\sigma,j_\sigma|\{n\})= \delta [n_{i_\sigma},0]\; \delta [n_{j_\sigma},1]\; d_{j_\sigma,i_\sigma}\;,\\
S(i_\sigma,j_{-\sigma}|\{n\})= \delta [n_{i_\sigma},0]\; \delta [n_{j_{-\sigma}},1] \;w_{j_{-\sigma},i_\sigma} \;.
\end{eqnarray}

The first term in Eq.(5) describes the rate at which
a given distribution decays due to electron tunneling to and off the
central electrode.
The second and third (fourth and fifth) terms, on the other hand, describe
the rate at which the probability of a given distribution increases due to
tunneling processes from (to)
the central electrode to (from) the external ones.
The terms with the coefficients $H$, $D$ and $S$ describe the electronic
(spin-conserving)
and spin-flip relaxation processes inside the central electrode.
The transition probability from the level $E_{i_\sigma}$ to the level
$E_{j_\sigma}$ is $d_{i_\sigma,j_\sigma}$ and to the level $E_{j_{-\sigma}}$  is
$w_{i_\sigma,j_{-\sigma}}$.
The master equation (\ref{4}) has a  general form which  includes internal
relaxation processes on the island and also the influence
of gate voltage $V_G$.
In Eq.(5) we assumed that electrons
in the source and drain
electrodes are in thermal equilibrium, while the
electrons in the island can be generally out of equilibrium.

It is convenient to define
the probability $P(N_{\uparrow},N_{\downarrow})$, that the island is occupied
by $N_{\uparrow}$ electrons with spin $\sigma=\uparrow$
and $N_{\downarrow}$ electrons with spin $\sigma=\downarrow$, respectively.
\begin{equation}\label{6}
P(N_{\uparrow},N_{\downarrow})= \sum_{\{n\}}
P(\{n\}) \delta [N_{\uparrow},\sum_{i_\uparrow}n_{i_\uparrow}]
\;\delta [N_{\downarrow},\sum_{i_\downarrow} n_{i_\downarrow}]\;.
\end{equation}
>From Eq.(\ref{4}) one finds the following master equation for
$P(N_{\uparrow},N_{\downarrow})$ in the stationary state:
\begin{eqnarray}\label{7}
0=\frac{\partial P(N_{\uparrow},N_{\downarrow})}{\partial t}=
-P(N_{\uparrow},N_{\downarrow})[A(N_{\uparrow},N_{\downarrow})
+H( N_{\uparrow},N_{\downarrow})]\nonumber\\+P(N_{\uparrow}+1,N_{\downarrow})
B_{\uparrow}(N_{\uparrow}+1,N_{\downarrow}) +P(N_{\uparrow},N_{\downarrow}+1)
B_{\downarrow}(N_{\uparrow},N_{\downarrow}+1)\nonumber\\
+ P(N_{\uparrow}-1,N_{\downarrow})C_{\uparrow}(N_{\uparrow}-1,N_{\downarrow})
+ P(N_{\uparrow},N_{\downarrow}-1)C_{\downarrow}(N_{\uparrow},N_{\downarrow}-1)\nonumber\\
+P(N_{\uparrow}+1,N_{\downarrow}-1)S_{\uparrow,\downarrow}(N_{\uparrow},N_{\downarrow})
+P(N_{\uparrow}-1,N_{\downarrow}+1)S_{\downarrow,\uparrow}(N_{\uparrow},N_{\downarrow})\;.
\end{eqnarray}
We have defined here the following parameters:
\begin{eqnarray}\label{8}
A(N_{\uparrow},N_{\downarrow})=\sum_{\sigma}\sum_{i_\sigma} [1-F(E_{i_\sigma}|N_{\uparrow},
N_{\downarrow})]
\{\Gamma^l_{i_\sigma} f(E_{i_\sigma}+E^{l+}_{N^\star}-E_F)
+\Gamma^r_{i_\sigma} f(E_{i_\sigma}+E^{r+}_{N^\star}-E_F) \} \nonumber\\
+F(E_{i_\sigma}|N_{\uparrow},N_{\downarrow})
\{\Gamma^l_{i_\sigma} [1-f(E_{i_\sigma}+E^{l-}_{N^\star}-E_F)]
+ \Gamma^r_{i_\sigma}[1-f(E_{i_\sigma}+E^{r-}_{N^\star}-E_F)] \} \;, \\
B_{\sigma}(N_{\uparrow},N_{\downarrow})=
\sum_{i_\sigma}F(E_{i_\sigma}|N_{\uparrow},N_{\downarrow})
\{\Gamma^l_{i_\sigma} [1-f(E_{i\sigma}+E^{l-}_{N^\star}-E_F)] \nonumber \\
+\Gamma^r_{i_\sigma}[1-f(E_{i_\sigma}+E^{r-}_{N^\star}-E_F)] \} \;, \\
C_{\sigma}(N_{\uparrow},N_{\downarrow})= \sum_{i_\sigma}[1-F(E_{i_\sigma}|N_{\uparrow},
N_{\downarrow})]
\{\Gamma^l_{i_\sigma} f(E_{i_\sigma}+E^{l+}_{N^\star}-E_F)
+\Gamma^r_{i_\sigma} f(E_{i_\sigma}+E^{r+}_{N^\star}-E_F) \}\;,\\
H(N_{\uparrow},N_{\downarrow})=\sum_{\sigma}\sum_{i_\sigma}
\sum_{j_{-\sigma}}
[1-F(E_{i_\sigma}|N_{\uparrow},N_{\downarrow})] F(E_{j_{-\sigma}}|N_{\uparrow},N_{\downarrow})
 w_{j_{-\sigma},i_{\sigma}} \;, \\
S_{\sigma,-\sigma}(N_{\uparrow},N_{\downarrow})=\sum_{j_{-\sigma}}
\sum_{i_\sigma}
F(E_{i_\sigma}|N_{\uparrow},N_{\downarrow})[1- F(E_{j_{-\sigma}}|N_{\uparrow},N_{\downarrow})]
 w_{i_\sigma,j_{-\sigma}} \;,
\end{eqnarray}
and also have introduced the function
\begin{equation}\label{9}
F(E_{i_\sigma}|N_{\uparrow},N_{\downarrow})=
\frac{1}{P(N_{\uparrow},N_{\downarrow})} \sum_{\{n\}} P(\{n\}) \delta [n_{i_\sigma},1]\;
\delta [N_{\uparrow},\sum_{i_\uparrow}n_{i_\uparrow}] \;
\delta [N_{\downarrow},\sum_{j_\downarrow} n_{j_\downarrow}],
\end{equation}
which is the probability that the level $E_{i_\sigma}$
is occupied when the island contains
$N_{\uparrow}$ electrons of spin $\sigma =\uparrow$
and $N_{\downarrow}$ electrons of spin $\sigma =\downarrow$. Note, that in Eqs (14) to (16)
$N^\star$ is the number of excess electrons on the island corresponding to the numbers
$N_\uparrow$ and $N_\downarrow$. When either $N_\uparrow$ or $N_\downarrow$
increases (decreases) by one, the corresponding number $N^\star$ also increases
(decreases) by one.

One can easily show that the following relations are fulfilled:
\begin{eqnarray}\label{9a}
1-F(E_{i_\sigma}|N_{\uparrow},N_{\downarrow}) = \frac{1}{P(N_{\uparrow},N_{\downarrow})}
 \sum_{\{n\}} P(\{n\})
\delta [n_{i_\sigma},0]\;\delta [N_{\uparrow},\sum_{i_\uparrow}n_{i_\uparrow}]
\delta [N_{\downarrow},\sum_{j_\downarrow} n_{j_\downarrow}] \;, \\
F(E_{j_{-\sigma}}|N_{\uparrow},N_{\downarrow})[1-F(E_{i_\sigma}|N_{\uparrow},N_{\downarrow})]
= \frac{1}{P(N_{\uparrow},N_{\downarrow})} \sum_{\{n\}} P(\{n\}) \nonumber \\
\times\delta [n_{i_\sigma},0]\;\delta [n_{j_{-\sigma}},1]\;\delta [N_{\uparrow},
\sum_{i_\uparrow}n_{i_\uparrow}]
\delta [N_{\downarrow},\sum_{j_\downarrow} n_{j_\downarrow}]\;.
\end{eqnarray}

In the following we restrict ourselves to the case of short electronic
(spin-conserving) relaxation time, $d_{i_\sigma,j_\sigma} \gg
\Gamma^r_{i_\sigma} ,\Gamma^l_{i_\sigma}$,
while the spin relaxation
time is much longer,
$d_{i_\sigma,j_\sigma} \gg \Gamma^r_{i_\sigma} ,\Gamma^l_{i_\sigma} \gg
w_{i_\uparrow,j_\downarrow}$.
The fast electronic  relaxation leads to
thermalization of electrons with a given spin orientation.
The two spin subsystems, however, are not in equilibrium and correspond
to different chemical potentials $\mu_\uparrow$ and $\mu_\downarrow$,
which are determined by $N_\uparrow$ and $N_\downarrow$, respectively.

The free energy of internal degrees of freedom can be expressed as
\begin{eqnarray}\label{15}
{\cal F}(N_{\uparrow},N_{\downarrow})=-k_BT \sum_{\sigma}\ln \left[ \sum_{\{n\}}
\delta [N_{\sigma},\sum_{i_\sigma}n_{i_\sigma}]
\exp \left( -\frac{1}{k_BT} \sum_{i_\sigma} E_{i_\sigma}n_{i_\sigma} \right)
 \right]\;
\end{eqnarray}
and the probability $F(E_{i\sigma}|N_{\uparrow},N_{\downarrow})$ is then given
by the following expression
\begin{eqnarray}\label{14}
F(E_{i_\sigma}|N_{\uparrow},N_{\downarrow})=
\exp \left(\frac{{\cal F}(N_{\uparrow},N_{\downarrow})}{k_BT} \right) \sum_{\{n\}}
\delta [n_{i_\sigma},1]\;\delta  [N_{\uparrow},\sum_{i_\uparrow}n_{i_\uparrow}] \nonumber \\
\times\delta [N_\downarrow,\sum_{i_\downarrow}n_{i_\downarrow}]
\exp\left( -\frac{1}{k_BT} \sum_{\sigma}\sum_{i_\sigma} E_{i_\sigma}n_{i_\sigma} \right).
\end{eqnarray}

In the limit $k_BT \gg \Delta E$ the distribution
function $F$ can be approximated by the Fermi-Dirac distribution
\begin{equation}\label{20}
F(E_{i_\sigma}|N_{\uparrow},N_{\downarrow})=f(E_{i_\sigma}-\mu_\sigma
(N_{\sigma}))\;,
\end{equation}
where the chemical potential $\mu_\sigma (N_{\sigma})$ is to be determined from the equation
\begin{equation}\label{21}
\sum_{i_\sigma}f(E_{i_\sigma} -\mu_\sigma (N_{\sigma}))=N_{\sigma}.
\end{equation}
In the regime $k_BT \le \Delta E$, the distribution function for only two levels
is significantly different from zero or one, so one may treat the system
as effectively a two-level one.$^{10}$ If we denote the relevant energy levels as
$E_{1_\sigma}$ and $E_{2_\sigma}$, then from the Gibbs
distribution one finds the following expression for the function $F$:
\begin{equation}\label{22}
F(E_{i_\sigma}|N_{\uparrow},N_{\downarrow})=\frac{\exp(-E_{i_\sigma}/k_BT )}
{\exp(-E_{1_\sigma}/k_BT) + \exp(-E_{2_\sigma}/k_BT)}= \left\{ 1+ \exp \left[
\frac{E_{i_\sigma}-\frac{1}{2}(E_{1_\sigma}+E_{2_\sigma})}{\frac{1}{2}k_BT} \right] \right\}^{-1}
\end{equation}
for $i_\sigma =1_\sigma$ and  $i_\sigma =2_\sigma$.

When we express electric current $I$ (see Eq.(1))
in terms of the distribution function $F$, then it is given by
\begin{eqnarray}\label{13}
I=e\sum_{N_{\uparrow},N_{\downarrow}}\sum_{\sigma}\sum_{i_\sigma}
P(N_{\uparrow},N_{
\downarrow}) \Bigl\{ [1-F(E_{i_\sigma}|N_{\uparrow},N_{\downarrow})]
\Gamma^r_{i_\sigma} f(E_{i_\sigma}+E^{r+}_{N^\star}-E_F) \nonumber\\
- F(E_{i_\sigma}|N_{\uparrow},N_{\downarrow}) \Gamma^r_{i\sigma}
[1-f(E_{i_\sigma}+E^{r-}_{N^\star}-E_F)] \Bigr\}.
\end{eqnarray}

The assumption of thermal equilibrium (for a particular spin orientation)
on the central electrode requires $\tau_{in} \le\tau_I$, where
$\tau_I=e/I$ is the injection time and $\tau_{in}$ is the inelastic
relaxation time.$^{10}$
Thus, our further analysis is valid when   $\tau_{in} \le\tau_I$.
At low temperatures the main contribution to $\tau_{in}$ is due to electron-electron
and electron-phonon interactions. Experimentally, $\tau_{in}$ was extensively
studied in the past by means of the weak localisation phenomenon and typical values
of $\tau_{in}$ were found to be between $10^{-12}$ and $10^{-10}$ sec.$^{17-19}$
In small clusters the relaxation time $\tau_{in}$ can be larger than its corresponding
bulk value.$^{13}$

\section{CHARACTERISTICS OF FERROMAGNETIC SET'S}

In this section we describe numerical results obtained on
basic characteristics of the junction. To simplify the picture arising
from discretization of energy levels of the island, we assume that the levels
are spin degenerate (nonmagnetic island) and equally separated with the
inter-level spacing $\Delta E$. In that case the numbers $N_{0\uparrow}$ and
$N_{0\downarrow}$ are equal, so that the excess magnetic moment
is equal to the total magnetic moment, i.e.,  $M^\star =M$.
Assuming additionally that the density of states $D^{l(r)}_\sigma$
in external electrodes and the matrix elements $M^{l(r)}_{i_\sigma}$
are independent of energy ($M^{l(r)}_{i_\sigma}$=$M^{l(r)}_{\sigma}$),
one can rewrite Eq.(27) as
\begin{eqnarray}\label{13}
I=\sum_\sigma {\Delta E\over e^2R^r_\sigma}\sum_{N_{\uparrow},N_{\downarrow}}\sum_{i_\sigma}P(N_{\uparrow},N_{
\downarrow}) \Bigl\{ [1-F(E_{i_\sigma}|N_{\uparrow},N_{\downarrow})]
f(E_{i\sigma}+E^{r+}_{N^\star}-E_F) \nonumber\\
- F(E_{i_\sigma}|N_{\uparrow},N_{\downarrow}) [1-f(E_{i\sigma}+E^{r-}_{N^\star}-E_F)] \Bigr\} \;,
\end{eqnarray}
where $R^r_\sigma$ is the resistance of the right junction,
$(R^r_\sigma)^{-1}=(2\pi / \hbar )\vert M^{r}_{\sigma}\vert^2D^{r}_{\sigma}(1/\Delta E)$.
Introducing in a similar way also the resistance $R^l_\sigma$ of the left junction,
one can express the parameters (14) to (18) in terms of $R^r_\sigma$ and
$R^l_\sigma$ and then calculate the probability $P(N_\uparrow ,N_\downarrow )$
from the master equation (13).

The formalism described above makes use of the two-dimensional
space of states $(N_{\uparrow},N_{\downarrow})$, in contrast to
the spinless case, where the relevant space is one-dimensional.
Basic physical characteristics of the system are then determined
by the probability $P(N_{\uparrow},N_{\downarrow})$ introduced in
Eq.(11). When expressed in terms of $N_{\uparrow}^\star$ and
$N_{\downarrow}^\star$, this probability will be denoted as
$P^\star (N_{\uparrow}^\star,N_{\downarrow}^\star)$. In the
$(N_{\uparrow}^\star ,N_{\downarrow}^\star )$ space this
probability is localized on a small number of points, as shown
in Fig.2 for a few values of the bias voltage in
the parallel and antiparallel configurations. The area of the black dots
located at the points in the ($N^\star_\uparrow ,N^\star_\downarrow$) space
is proportional to the corresponding probabilty
$P^\star (N_{\uparrow}^\star ,N_{\downarrow}^\star )$. In each case the total area
of all black dots is normalized to unity. For $V=5$ mV there is no excess electron
on the island ($N_{\uparrow}^\star =N_{\downarrow}^\star =0$), as this value of $V$ is within the
Coulomb blockade region (see Fig.3a, where the corresponding $I-V$  curves are shown).
For $V=20$ mV (within the first plateau above the threshold voltage in the $I-V$ curves)
the points with dominant
probability $P^\star (N_{\uparrow}^\star ,N_{\downarrow}^\star )$
 are located on the line corresponding to
$N_{\uparrow}^\star +N_{\downarrow}^\star =N^\star=1$. There are also points corresponding
to $N^\star=0$, but the corresponding probabilty is significantly smaller and these points
 will be neglected in the further discussion. It is interesting to note that
different points correspond to different values of the excess spin on the island.
In the parallel configuration these points are distributed
symmetrically on both sides of the line
corresponding to $N^\star_\uparrow =N^\star_\downarrow$. Consequently, the
average spin accumulated on the island is zero, contrary to the
antiparallel configuration, where the average spin accumulated on the island is nonzero.
For $V=30$ mV the probability $P^\star (N_{\uparrow}^\star ,N_{\downarrow}^\star )$
is significant for $N^\star =1$ and $N^\star =2$. This value of $V$ corresponds to
the transition between the first and second steps in the $I-V$ curves.
As before, the average spin accumulated
on the island vanishes in the parallel configuration, whereas in the antiparallel
configuration it is different from zero.
Note, that the number of different values of the excess spin on the island
is now smaller. The situation for $V=40$ mV is qualitatively
similar to that for $V=20$ mV, but the number of black dots is larger. Generally, one
can note from Fig.2, that
when the bias voltage $V$ increases, the localization area of the
probability $P^\star (N_{\uparrow}^\star ,N_{\downarrow}^\star )$
shifts to new stationary points and embraces more and more points
in the $(N_{\uparrow}^\star ,N_{\downarrow}^\star )$ space.
Similar tendency can be  observed when the temperature increases.
Therefore, in order to get convergence
in numerical calculations, the number of states taken into account
was dynamically changed with increasing bias voltage and temperature.

\subsection{Bias voltage characteristics}

Figure 3a shows the current-voltage characteristics of a junction
with a nonmagnetic island and ferromagnetic source and sink
electrodes. The single-junction resistances in the parallel
configuration have been assumed to be
$R_{l\uparrow}=200\;$M$\Omega$, $R_{l\downarrow}=100\;$M$\Omega$
for the left junction and $R_{r\uparrow}=4\;$M$\Omega$,
$R_{r\downarrow}=2\;$M$\Omega$ for the right one. In the
antiparallel configuration  the magnetization of the right
electrode is reversed and the corresponding resistances are
$R_{r\uparrow}=2\;$M$\Omega$ and $R_{r\downarrow}=4\;$M$\Omega$.
Note, that the same spin asymmetry factor
$p_j=R_{j\downarrow}/R_{j\uparrow}$ has been assumed for both
junctions in the parallel configuration, $p_l=p_r=1/2$. Owing to a
large difference between  the resistances of the left and right
junctions, the Coulomb steps in the $I$-$V$ characteristics are
clearly seen.$^{11,17}$  Since $C_l > C_r$ in the case
considered here, the
threshold voltage $V_{\rm th}$, below which the current is blocked
($N^\star =0$), is approximately equal to $V_{\rm th}=(C/C_l)(2E_c+\Delta
E)/2e\approx 10.2$mV. The large steps in Fig.3a correspond
respectively to $N^\star = 1$, $2$, ..., and their length is
$V_p\approx (C/C_l)(4E_c+\Delta E)/2e\approx 19.1$ mV. There are also
additional small steps of length $V_s=(C/C_l)(\Delta
E/e)\approx 2.7$ mV, which result from discretness of the energy
spectrum of the island and correspond to opening a tunneling
channel with a new value of the excess spin on the island (new
value of $M$). Position of the steps is clearly seen in the
$dI/dV$ curves shown in Fig.3b for the antiparallel configuration.
The large peaks correspond there to the Coulomb steps while the
small ones to the steps due to discrete energy spectrum. This
behavior is qualitatively similar to that observed experimentally in tunneling
through small Al particles$^{18}$ or through $C_{60}$ molecules.$^{19}$

The $I$-$V$ curves in the parallel and  antiparallel
configurations are different (solid and dashed curves in Fig.3a).
Consequently, the corresponding resistances of the whole system
are also different in both configurations; $R_{p}$ and  $R_{ap}$,
respectively. This, in turn, results in tunnel magnetoresistance
(TMR), which is described quantitatively by the ratio
$TMR=(R_{ap}-R_p)/R_{p}$.$^{20}$ The bias dependence of TMR is
shown in Fig.3c. As one can see, TMR oscillates with increasing $V$
with the period $V_p$. The amplitude of the oscillations decreases
with increasing voltage.  In the limit $V\gg E_c/e$ the system can
be treated as a set of ohmic resistors with the total resistance
$R^{-1}=(R_{l\uparrow}+R_{r\uparrow})^{-1}+(R_{l\downarrow}+
R_{r\downarrow})^{-1}$.
 In our case the total limiting resistances for the antiparallel and parallel
configurations are respectively $R_{ap}=68.65$ M$\Omega$ and
$R_p=68$ M$\Omega$, which gives the asymptotic value of TMR equal
approximately to 0.01. This value can be larger for systems with
either larger spin asymmetry in the single-junction resistances,
or smaller difference between the resistances of left and right
junctions.

For the parameters assumed in numerical calculations, the
incoming electrons pass
through the less resistive  and more capacitive junction, while the
outgoing electrons pass through the
more resistive and less capacitive one.
In that case electrons accumulate on the island when a bias voltage $V$ is applied.
Fig.4a presents
the bias dependence of the charge accumulation. The steps in the curves
show that the average charge $<Q>$ accumulated on the island is close to $1e$,
$2e$, ... and is almost constant between the steps. Plot of the root mean square,
rms$(N^\star )=[<N^{\star 2}>-<N^\star >^2]^{1/2}$,
as a function of $V$ is presented in Fig.4b.
The charge fluctuations are large at the steps, where a new charge channel becomes open,
i.e. when $N^\star \to N^\star +1$. Between the steps fluctuations are rather small.

When the right and left junctions correspond  to different spin asymmetry factors, then
not only charge but also spin is
 accumulated on the island. For the junction assumed in Fig.3 this happens in the
antiparallel configuration. The plot of $<M>$ as a function of $V$
is shown in Fig.5a. Indeed, there is almost no spin accumulation
in the parallel configuration, whereas a significant spin
accumulation occurs in the antiparallel configuration, which
varies oscillatory-like with increasing $V$. The origin of the
oscillatory behavior is described in Ref. [5], Here, we only note
that beginning from the threshold voltage, the average $<M>$
increases with increasing $V$ up to $<M>=3$, which occurs at
$V\approx 20$ mV. At this value of $V$ a new charge channel,
corresponding to $N^\star =2$, becomes open for one spin orientation,
which reduces spin accumulation. The average $<M>$ starts to
increase again at $V\approx 30$ mV, and the second oscillation
period in the spin accumulation begins.

Figure 5b shows fluctuations of the induced magnetic moment on the
island. Although there is almost no spin accumulation in the
parallel configuration, the curve representing spin fluctuations
in the parallel configuration is similar to that  for the
antiparallel one. Moreover, the fluctuations in $M$ are even
larger in the parallel configuration than in the antiparallel one,
because the space of states  available for fluctuations is reduced
by the spin accumulation. Numerical analysis of the probability
distribution $P^\star (N_{\uparrow}^\star ,N_{\downarrow}^\star )$
on the first Coulomb step in the parallel configuration shows two
high maxima corresponding to opposite induced magnetic moments
(which gives $<M>=0$). The separation between the maxima increases
with increasing voltage and then decreases when $V$ exceeds  23 mV.
Behavior of the spin fluctuation with increasing bias voltage
resembles behavior of the average spin accumulated on the island.
The fluctuations vary oscillatory-like with increasing V, with the
same phase and period as the oscillations in $<M>$. It is also
interesting to note, that every second peak of the spin
accumulation and spin fluctuation in Fig.5a and Fig.5b have a
similar shape. This additional periodicity is due to variation of
the ground state from the state with odd number of electrons to
that with even number of electrons on the island (the
$(N_{\uparrow}^\star ,N_{\downarrow}^\star )$ space is different
along the diagonal corresponding to $N^\star$ odd or $N^\star$
even).

The maximum current in our numerical results is of the order of 1 nA,
which corresponds to the lowest value of the injection time,
$\tau_I\approx 5\times 10^{-10}$ sec.
Thus, the numerical results are valid in the whole range of applied voltage when
$\tau_{in}\le 5\times 10^{-10}$ sec, which can be obeyed in real systems
at $T\ge T_l$, where $T_l$ is of the order of 1 K.
This estimate is consistent with that in Ref.[13], where the inelastic relaxation time
at 30 mK was estimated for a small Al cluster. When adapted to our
value of $\Delta E$, this estimate gives $\tau_{in}$ of the order
of $10^{-9}$ sec. Since
$\tau_{in}$ decreases with increasing temperature, $T_l=1$ K as
the lower limit  for validity of our numerical calculations seems to be quite
reasonable. This temperature is low enough to observe the level quantization.
The lowest temperature assumed in our numerical calculations is 2.3 K, which
is above the lower limit $T_l$ and also sufficiently below the upper limit,
determined by the condition $k_BT\approx \Delta E$,
above which the quantization effects disappear. It is
also worth to note, that $\tau_I$ can be made longer by an increase in the
junction resistances. Thus, for realistic $\tau_{in}$ one can always find
a range of parameters, where our description is valid.

\subsection{Temperature dependence}

The numerical results presented above were calculated for $k_BT$  much
smaller than the charging energy
$E_c$ and also smaller than the level spacing $\Delta E$. The two energy scales were then
clearly seen in all characteristics of the system. When the temperature increases
the probability $P^\star (N_{\uparrow}^\star ,N_{\downarrow}^\star )$
spreads over larger area in the
$(N_{\uparrow}^\star ,N_{\downarrow}^\star )$ space
and the peaks become smaller.
This has a significant influence on transport properties.

In Figs. 6a and 6b we show the current characteristics (a) and TMR
(b) for different temperatures. The Coulomb steps in the $I$-$V$
curves disappear at high temperatures, and the current becomes
ohmic with the classical value of the resistance,
$R^{-1}=(R_{l\uparrow}+R_{r\uparrow})^{-1}+(R_{l\downarrow}+R_{r\downarrow})^{-1}$.
The corresponding value of TMR is then equal to about $0.01$ and
is almost voltage independent. The small steps in the $I-V$ curves
and TMR, which result from discreteness of the electronic
structure of the island, disappear rather quickly with increasing
temperature, much earlier than the Coulomb steps do. For $T=11.6$
K (the thermal energy is equal to 1~meV and is three times smaller
than $\Delta E=3$~meV) most of the small steps disappear, but
there are still well defined Coulomb steps and large oscillations
due to charging effects (in this case $E_c=6.02$~meV).

The influence of increasing temperature on the bias dependence of
charge accumulation and charge fluctuations is shown in Fig.7a.
and 7b, respectively. The curves representing charge accumulation
at different temperatures are similar to the corresponding $I$-$V$
characteristics. The effects due to discreteness of energy levels
and due to discrete charging  gradually disappear with increasing
temperature. At $T=58$ K the charge accumulation becomes a linear
function of $V$, as it should be in an ohmic system. Oscillations
in the charge fluctuations are less sensitive to the temperature.
As follows from Fig.7b, they are periodic functions of $V$ and the
periodicity survives even at $T=58$ K, where the $I-V$ curves have
already ohmic character.

Spin accumulation and spin fluctuations at different temperatures
are shown in Fig.8. The oscillations with increasing bias voltage
disappear when the temperature increases, quite similarly as the
oscillations in charge accumulation and charge fluctuations
(Fig.7). At $T=58$ K the spin accumulation and spin fluctuations
vary almost linearly with increasing bias.

\subsection{Gate voltage dependence}

Consider now transport characteristics of the system as a function
of the gate voltage $V_g$, which is related to the induced charge
$Q_{\rm in}=V_gC_g$ on the island. Assume a constant bias voltage
which is above the threshold voltage and corresponds to the plateau
between the first and second Coulomb steps, say $V=15$ mV. Figures
9a and 9b show the $I-V$ characteristics and TMR, respectively, as
a function of the gate voltage, and calculated for  $T=2.3$ K,
$T=11.6$ K and $T=34.8$ K. Figure 10, on the other hand, shows
charge (a) and spin (b) accumulation on the island, calculated for
the same temperatures as in Fig.9. Electric current, TMR and spin
accumulation are periodic functions of $V_g$, with the period
$V_g^p=2e/C_g\approx 107$ mV corresponding to $\Delta Q_{\rm
in}=2e$. The curve corresponding to charge accumulation is similar to
that representing charge accumulation as a function of the bias
voltage (Fig.4a). Due to asymmetry of the states with odd and even
numbers of electrons on the island, the period $V_g^p$ is twice as
long as in the spinless case. At low temperatures the difference
between states with $N^\star $ odd and even is clearly seen in all
characteristics. At high temperatures, however, the difference
between those two cases disappears and period becomes the same as
in the splinless case. To understand this difference let us
analyze the situation in more details.
For $V=15$ mV and $V_g=0$ the average excess charge on the island
is close to $1e$. The probability $P^\star (N_{\uparrow}^\star
,N_{\downarrow}^\star )$ has then large peaks for
$N^\star =N_{\uparrow}^\star +N_{\downarrow}^\star =1$. An increase in
$V_g$ leads at low temperatures to an almost linear decrease of
the current (Fig.9a), while the charge accumulated on the island
remains almost unchanged (it increases very slowly, see Fig.10a).
Origin of the decrease in electric current can be explained as
follows. When $V_g$ increases the position of the Fermi level of
the island shifts to lower energies. This effectively reduces the
number of energy levels from which electrons can tunnel through
the left junction. In our case this junction has much larger
resistance than the right one, and therefore it is just the
junction which determines electric current flowing through the
system. Thus, an increase in $V_g$ results in a decrease in
electric current. (Opposite behavior, when current increases with
increasing $V_g$ is also possible for other parameters.)

At $V_g\approx 38$ mV the relevant states are $(N_{\uparrow}^\star
,N_{\downarrow}^\star )=(1,0)$ and (0,1). In the antiparallel
configuration the probability for these states is $P^\star
(1,0)=0.76$ and $P^\star (0,1)=0.20$, whereas in the parallel
configuration $P^\star (1,0)=P^\star (0,1)=0.48$. $V_g=38$ mV is
already close to the value at which ground state with one electron
more on the island becomes energetically more convenient. A small
increase of $V_g$ to 45 mV leads then to a large increase in the
charge accumulation, $\Delta <Q>\approx 1e$. This also leads to a
rapid increase in the electric current, roughly to the value it
had at $V_g=0$. The current increases because from the transport
point of view the system returns to the situation at $V_g=0$
(without counting the discreteness of the energy spectrum). For
$V_g=45$ mV there is only one relevant state, i.e. the state (1,1)
with the probability $P^\star (1,1)=0.70$ in the antiparallel and
$P^\star (1,1)=0.75$ in the parallel configurations. The spin
accumulation reaches then minimum at this point  (see Fig.10b).

At $V_g\approx 97$ mV a new ground state is formed and the system
goes over from the state with even number of electrons on the
island to the state with odd number of electrons. Close to
$V_g=97$ mV the relevant state is at (1,1), but a small increase
of $V_g$ leads the system to a new stationary state, in which the
states (2,1) and (1,2) are more important. This transition is
different from the one  at $V_g\approx 38$ mV. This difference is
clearly seen in TMR, which in this range of $V_g$ has a large deep
and becomes negative.

At higher temperatures the difference between the situations with
odd and even numbers of excess electrons  on the island is not
visible (see the curves  corresponding to $T=11.6$ K and 34.8K in
Fig.9 and Fig.10). The thermal energy is then comparable to the
energy needed for opening a new spin channel, i.e. $k_BT\approx
\Delta E$. Two states in the $(N_{\uparrow}^\star
,N_{\downarrow}^\star)$-space, for which $\Delta M=\pm 2$, are
difficult to be distinguished. Therefore, the periodicity is then
as in the spinless case.

\section{SUMMARY AND CONCLUSIONS}

We have developed formalism for calculating electric current,
spin and charge accumulation and TMR in ferromagnetic
SET's with a small central electrode -- small enough so that the
discrete structure of its energy spectrum plays a significant role.
We found two different scales in all characteristics of the junction;
the shorter one related to the discreteness of energy spectrum and the
longer one related to
discrete charging of the island with single electrons.
The features due to
discrete energy levels can be seen at low temperatures and disappear
relatively quickly with increasing temperature; much faster than
the features due to discrete charging ($\Delta E<<E_c$ in our case).

The junction characteristics are periodic functions of the
bias and gate voltages. At low temperatures the periods are twice as
long as the corresponding ones at high
temperature. This is because at low temperatures the situations with even
and odd numbers of electrons on the island can be distinguished, while
at high temperatures this difference disappears.

We have also shown that spin fluctuations can be significantly
larger than the charge fluctuations. Such large spin fluctuations
can play a significant role in the current noise.$^{21}$
\acknowledgments
        The paper is supported by the Polish State Committee for Scientific Research
under the Project No. 2 P03B 075 14.

\newpage

{\bf References}
\vskip 0.5cm
\begin{enumerate}
\item {For a review see {\it Single Charge Tunneling},
  edited by H. Grabert and M.H. Devoret, NATO ASI
  Series vol 294 (Plenum Press, New York 1992).}
\item {J. Inoue and S. Maekawa, Phys. Rev. B {\bf 53}, R11927 (1996);
  L.F. Schelp, A. Fert, F. Fettar, P. Holody, S.F. Lee, J.L. Maurice,
  F. Petroff and A. Vaures,  Phys. Rev. B {\bf 56}, R5747  (1997) .}
\item {K. Ono, H. Shimada, S. Kobayashi and Y. Outuka,
      J. Phys. Soc. Japan {\bf 65}, 3449 (1996);
      K. Ono, H. Shimada and Y. Outuka, {\it ibid} {\bf 66}, 1261 (1997);
      G. Reiss, H. Vizelberg, M. Bertram, I. M\"onch and J. Schumann,
      Phys. Rev. B {\bf 58}, 8893 (1998).}
\item {J. Barna\'s and A. Fert, Phys. Rev. Lett. {\bf 80}, 1058 (1998).}
\item {J. Barna\'s and A. Fert, Europhys. Lett. {\bf 44}, 85 (1998);
      J. Magn. Magn. Mater. {\bf 192}, L 391 (1999).}
\item {S. Takahashi and S. Maekawa, Phys. Rev. Lett. {\bf 80}, 1758 (1998).}
\item {A. Brataas, Yu.V. Nazarov, J. Inoue and G.E.W. Bauer,
       European Phys. Journ. B {\bf 9}, 421 (1999); Phys. Rev. B {\bf 59}, 93 (1999);
       K. Majumdar and S. Hershfield, Phys. rev. B {\bf 57}, 11 521 (1998). }
\item {J. Martinek, J. Barnas, G. Michalek, B.R. Bulka and A. Fert,
      J. Magn. Magn. Mater. {\bf 207}, L 1 (1999).}
\item {D.V. Averin and K.K. Likharev, J. Low Temp. Phys.
     {\bf 62}, 345  (1986);  D.V. Averin and A.N. Korotkov,
     Zh. Eksp. Teor. Fiz. {\bf97}, 1661 (1990).}
\item {C.W.J. Beenakker, Phys. Rev. B {\bf 44}, 1646 (1991).}
\item {D.V. Averin, A.N. Korotkov and K.K. Likharev, Phys. Rev. B {\bf 44}, 6199 (1991).}
\item {D.C. Ralph, C.T. Black and M. Tinkham, Physica {\bf 218B}, 258 (1996).}
\item {O. Agam, N.S. Wingreen, B.L. Altshuler, D.C. Ralph and M. Tinkham,
        Phys. Rev. Lett. {\bf 78}, 1956 (1997).}
\item {for a review see G. Bergmann, Phys. Reports {\bf 107}, 1 (1984);
       {\it Electron-electron interactions in disordered systems}, ed. A.L. Efros and
       M. Pollak, (North-Holland, Amsterdam 1985.}
\item {M. Gijs, C. Van Haesendock and
      Y. Bruynseraede. J. Phys. F: Met. Phys. {\bf 16}, 1227  (1986).}
\item {S. Aryainejad, Phys. Rev. B {\bf 32}, 7155  (1985);
        A.C. Sacharoff and R.M. Westervelt, Phys. Rev. B {\bf 32}, 662  (1985).}
\item {M. Amman, R. Wilkins, E. Ben-Jacob, P.D. Maker and
  R.C. Jaklewic, Phys. Rev. B  {\bf 43}, 1146  (1991).}
\item {D.C. Ralph, C.T. Black and M. Tinkham, Phys. Rev. Lett. {\bf 74}, 3241 (1995).}
\item {D. Porath, Y. Levi, M. Torabiah, and O. Millo,
       Phys. Rev. B {\bf 56}, 9829  (1997).}
\item {J.S Moodera, L.R. Kinder, T.M. Wong and R. Meservey,
  Phys. Rev. Lett. {\bf 74}, 3273  (1995).}
\item {B.R. Bulka, J. Martinek, G. Michalek and J. Barna\'s,
      Phys. Rev. B {\bf 60}, 12246 (1999).}

\end{enumerate}

\newpage
\begin{figure}
%\vskip 14cm
 \epsfxsize=14cm
 \epsffile{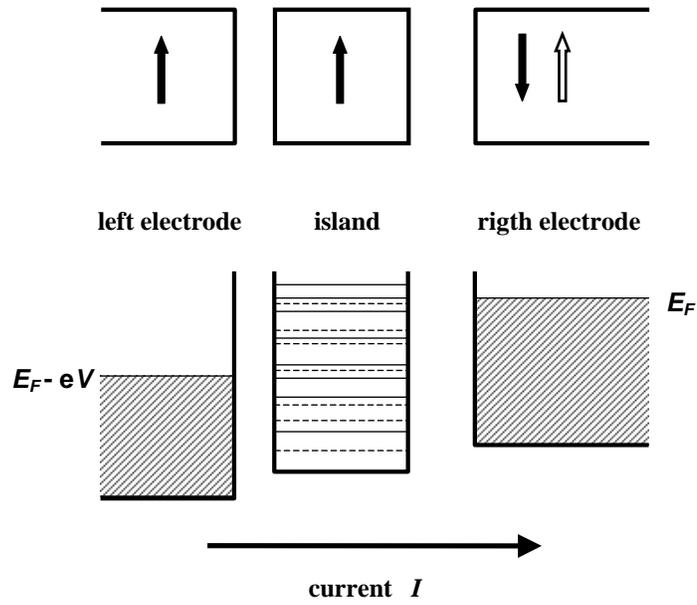 }
%\hskip 1cm
\caption {Geometry of the junction and schematic profile of the
potential energy when a bias voltage $V$ is applied. Discrete
energy levels of the island for both spin orientations are
indicated by the solid and dashed lines.} \label{Fig.1}
\end{figure}

\newpage
\begin{figure}[c]
%\vskip 14cm
\begin{center}
\epsfxsize=9cm
 \epsffile{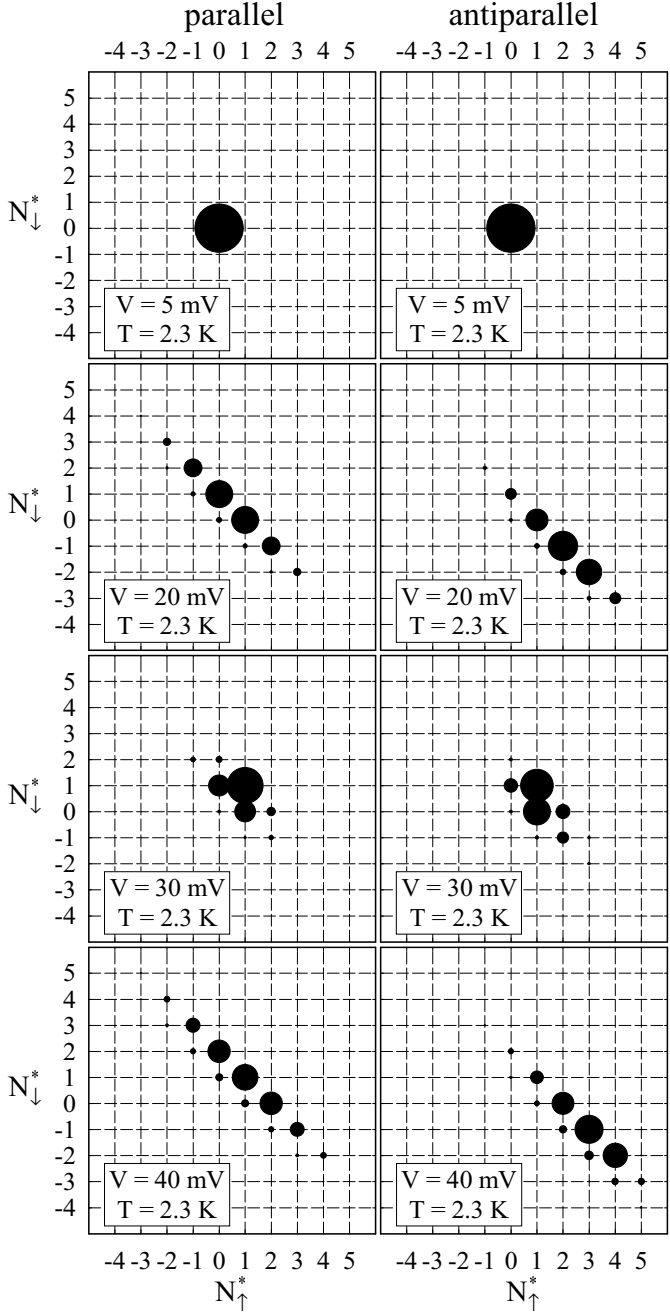 }
\end{center}
%\hskip 1cm
\caption
{Probability $P^\star (N^\star_\uparrow,N^\star_\downarrow)$ in
the $(N^\star_\uparrow,N^\star_\downarrow)$ space of states
(proportional to the area of the black dots) calculated for the
antiparallel and parallel configurations and for four different
values of the bias voltage. The junction resistances in the
parallel configuration are: $R_{l\uparrow}=200\;$M$\Omega$,
$R_{l\downarrow}=100\;$M$\Omega$, $R_{r\uparrow}=2\;$M$\Omega$ and
$R_{r\downarrow}=4\;$M$\Omega$, whereas in the antiparallel
configuration $R_{r\uparrow}=4\;$M$\Omega$,
$R_{r\downarrow}=2\;$M$\Omega$. The other parameters assumed in
numerical calculations are: $C_{l} =9\;$aF, $C_{r}=1.3\;$aF,
$C_{g}=3\;$aF, $E_{c}=6.02\;$meV, $\Delta E=1.8\;$meV, $V=26\;$mV,
$V_g=0$, $T=2.3$ K and $V_g=0$.} \label{Fig.2}
\end{figure}

\newpage
\begin{figure}
%\vskip 14cm
\epsfxsize=12cm
\epsffile{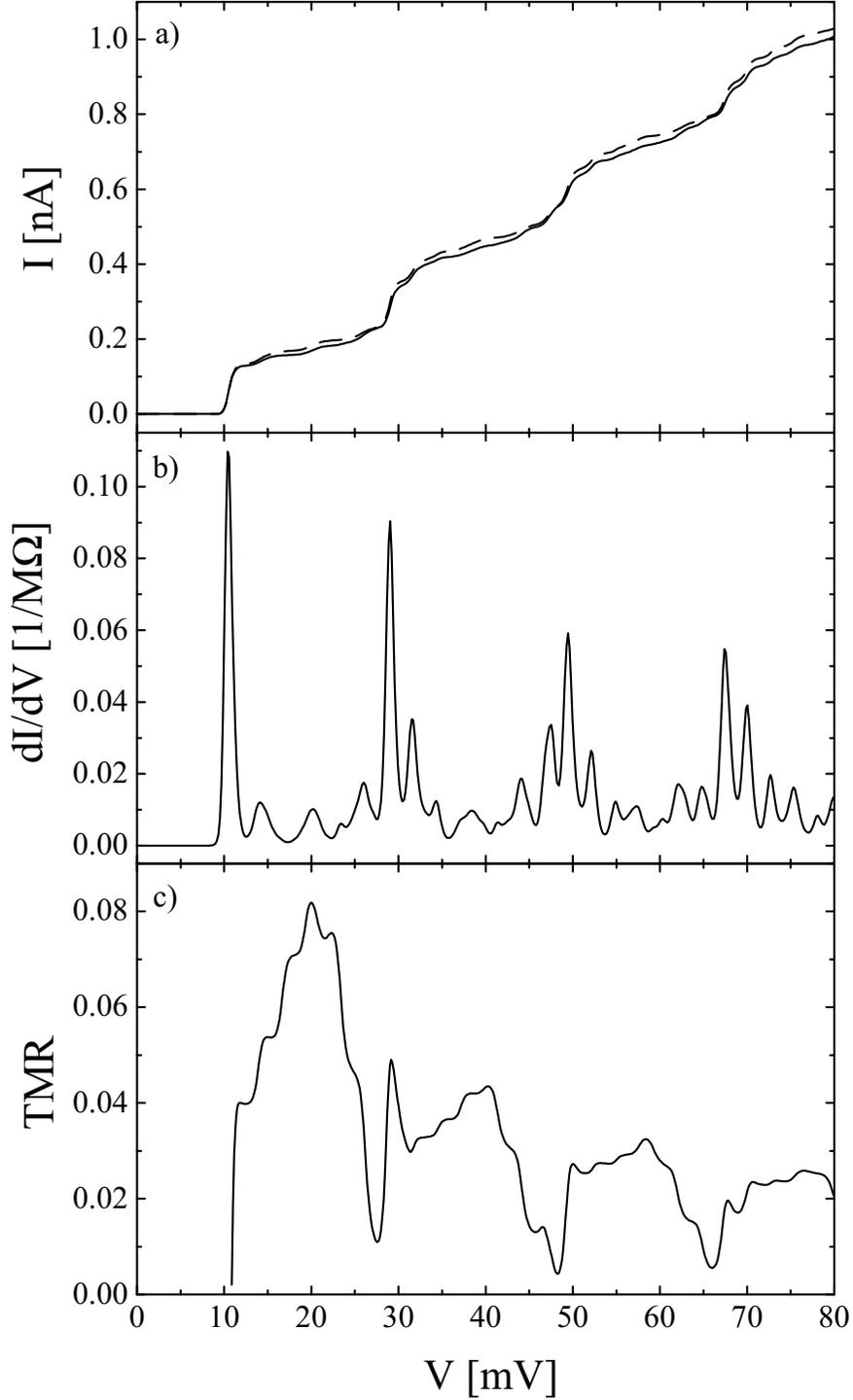 }
%\hskip 1cm
\caption{Voltage
dependence of the tunnel current $I$ (a), derivative $dI/dV$ (b),
and tunnel magnetoresistance (c) determined at $T=2.3\;K$. The
solid and dashed curves in (a) correspond to the antiparallel and
parallel configurations, respectively, whereas the plot in (b) is
for the antiparallel configuration only. The parameters of the
system are the same as in Fig.2.} \label{Fig.3}
\end{figure}

\newpage
\begin{figure}
%\vskip 14cm
%\epsfxsize=14cm
\epsffile{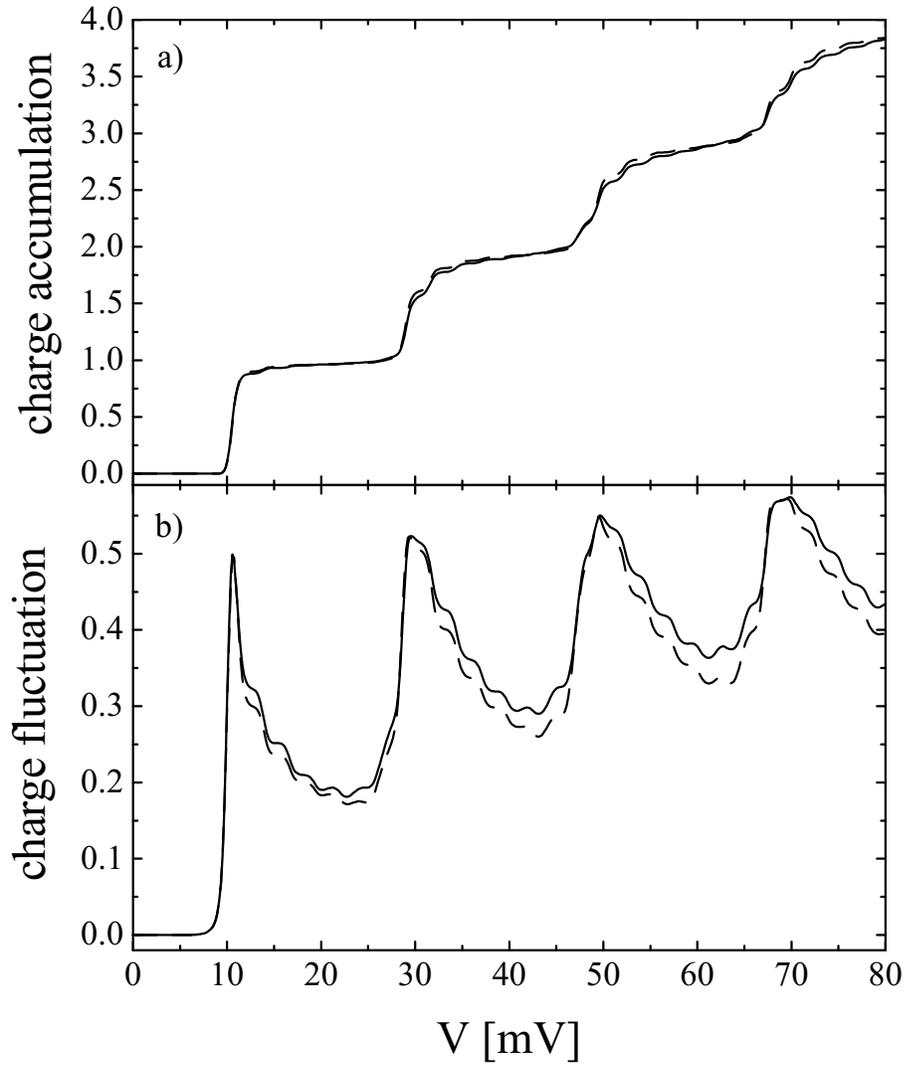 }
 \vspace{1cm}%\hskip 1cm
\caption{Charge accumulation $<N^\star >$ (a) and charge
fluctuations $[<N^{\star 2}-<N^\star >^2]^{1/2}$ (b) as a function
of the bias voltage. The solid and dashed curves corresponds to
the antiparallel and parallel configuration, respectively. The
parameters are the same as in Fig.2.} \label{Fig.4}
\end{figure}

\newpage
\begin{figure}
%\vskip 14cm \epsfxsize=14cm
\epsffile{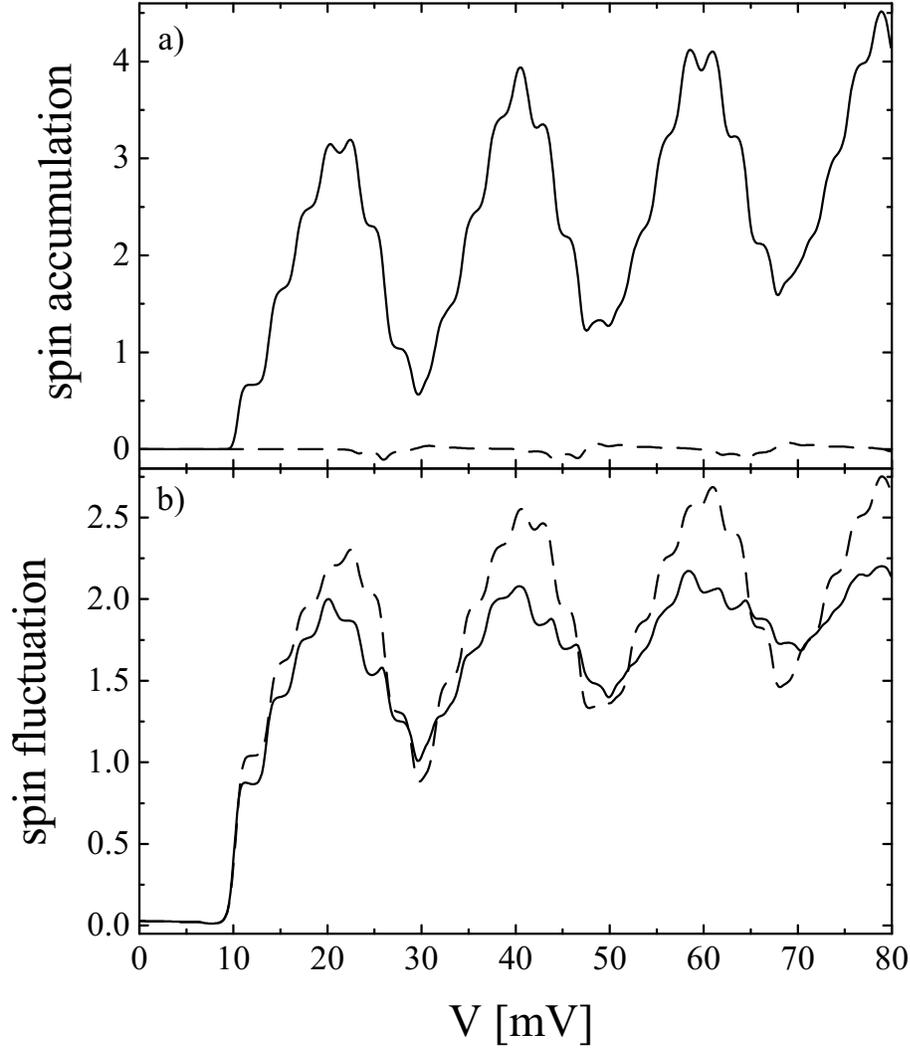 }
 \vspace{1cm}%\hskip 1cm
\caption{Spin accumulation $<M>=<N_\uparrow-N_\downarrow>$ (a) and
the spin fluctuations  $[<M^2-<M>^2]^{1/2}$ (b) as a function of
$V$ in the system defined in Fig.2. Solid and dashed curves are
for the antiparallel and parallel configuration, respectively.}
\label{Fig.5}
\end{figure}

\newpage
\begin{figure}
%\vskip 14cm \epsfxsize=14cm
\epsffile{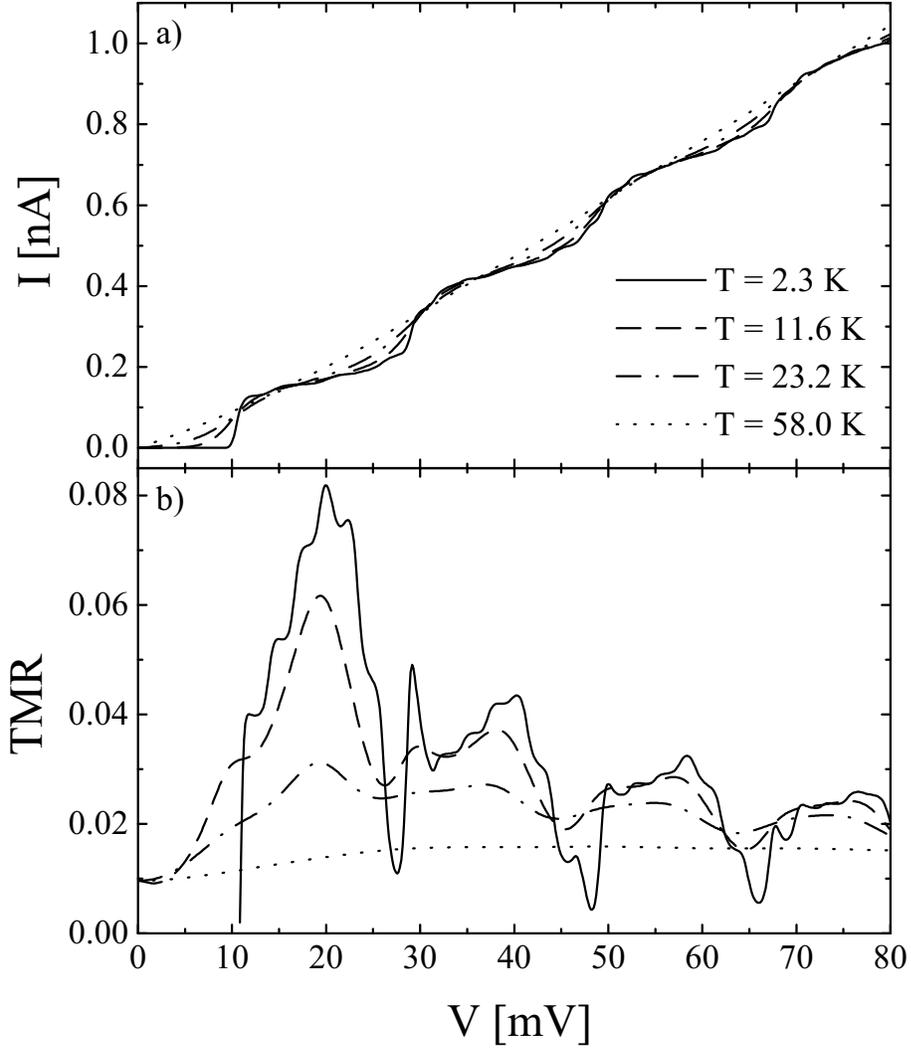 }
 \vspace{1cm}%\hskip 1cm
\caption{Bias dependence of
electric current in the antiparallel configuration (a)  and TMR
(b) for different temperatures and for $\Delta E/k_B=34.8$~K  and
$E_c/k_B=69.9$~K. The other parameters are the same as in Fig.2.}
\label{Fig.6}
\end{figure}

\newpage
\begin{figure}
%\vskip 14cm \epsfxsize=14cm
\epsffile{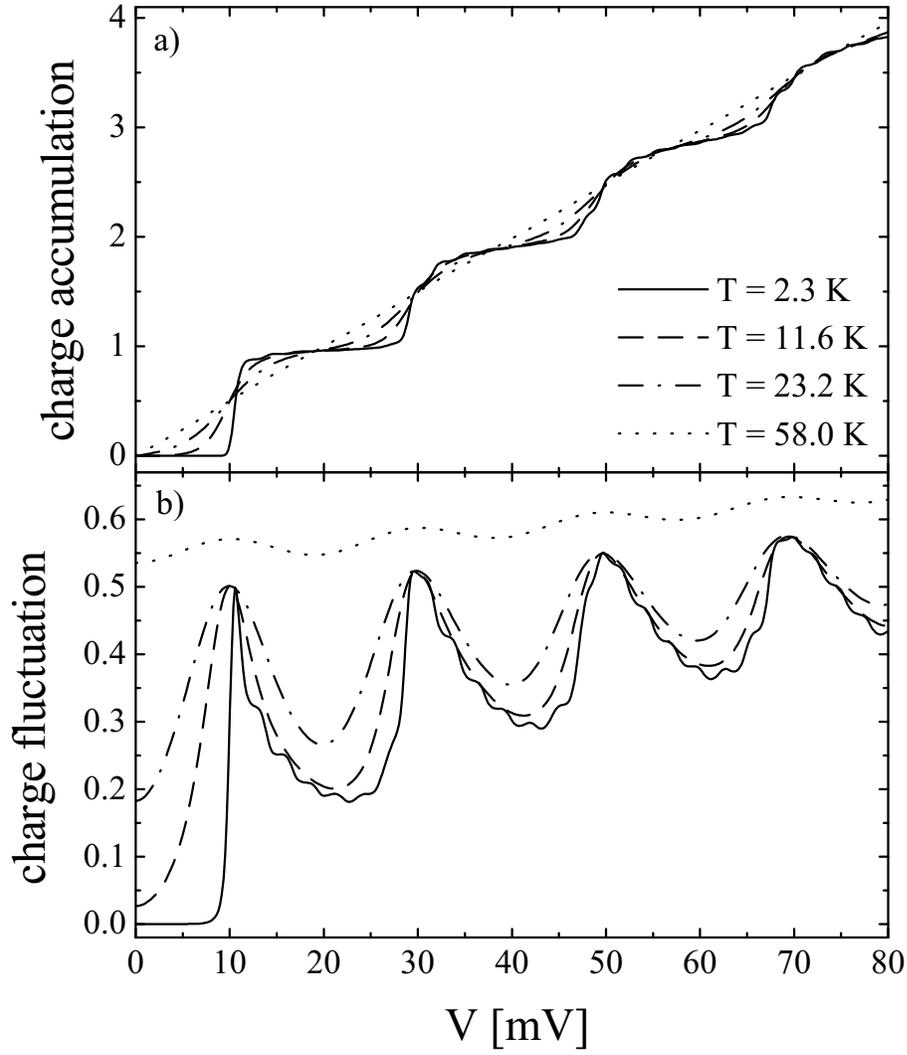 }
 \vspace{1cm}%\hskip 1cm
\caption{Bias dependence of the charge accumulation (a) and charge
fluctuations for different temperatures. The parameters are the
same as in Fig.2.} \label{Fig.7}
\end{figure}

\newpage
\begin{figure}
%\vskip 14cm \epsfxsize=14cm
\epsffile{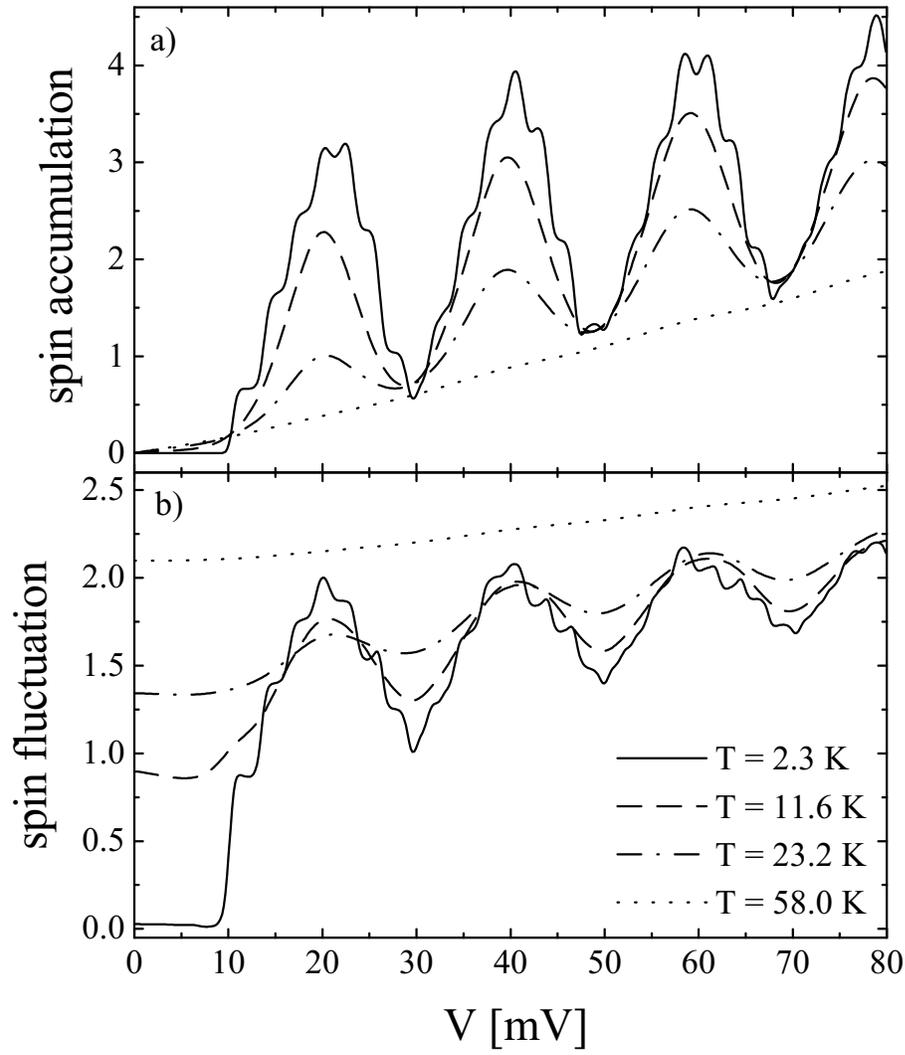 }
 \vspace{1cm}%\hskip 1cm
\caption{Voltage dependence of the
spin accumulation (a) and  spin fluctuations for different
temperatures. The parameters are the same as in Fig.2.}
\label{Fig.8}
\end{figure}

\newpage
\begin{figure}
%\vskip 14cm
%\epsfxsize=14cm
\epsffile{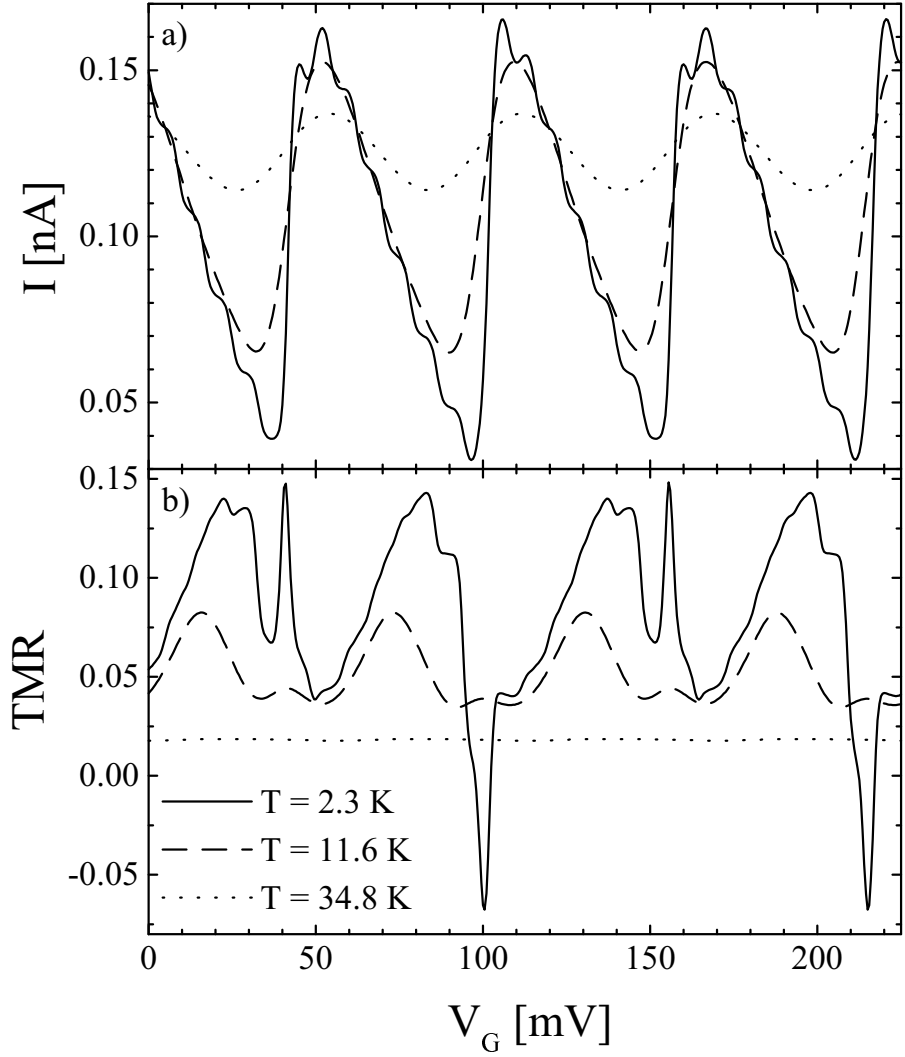 }
 \vspace{1cm}%\hskip 1cm
\caption{The tunneling current $I$ in the antiparallel
configuration (a) and TMR (b) as a function of the gate voltage
$V_g$ calculated for $V=15\;$mV and for $T=2.3$ K (solid curve),
$T=11.6$ K (dashed curve), $T=34.8$ K (dotted curve). The other
parameters of the system are as in Fig.2.} \label{Fig.9}
\end{figure}

\newpage
\begin{figure}
%\vskip 14cm
\epsfxsize=14cm
\epsffile{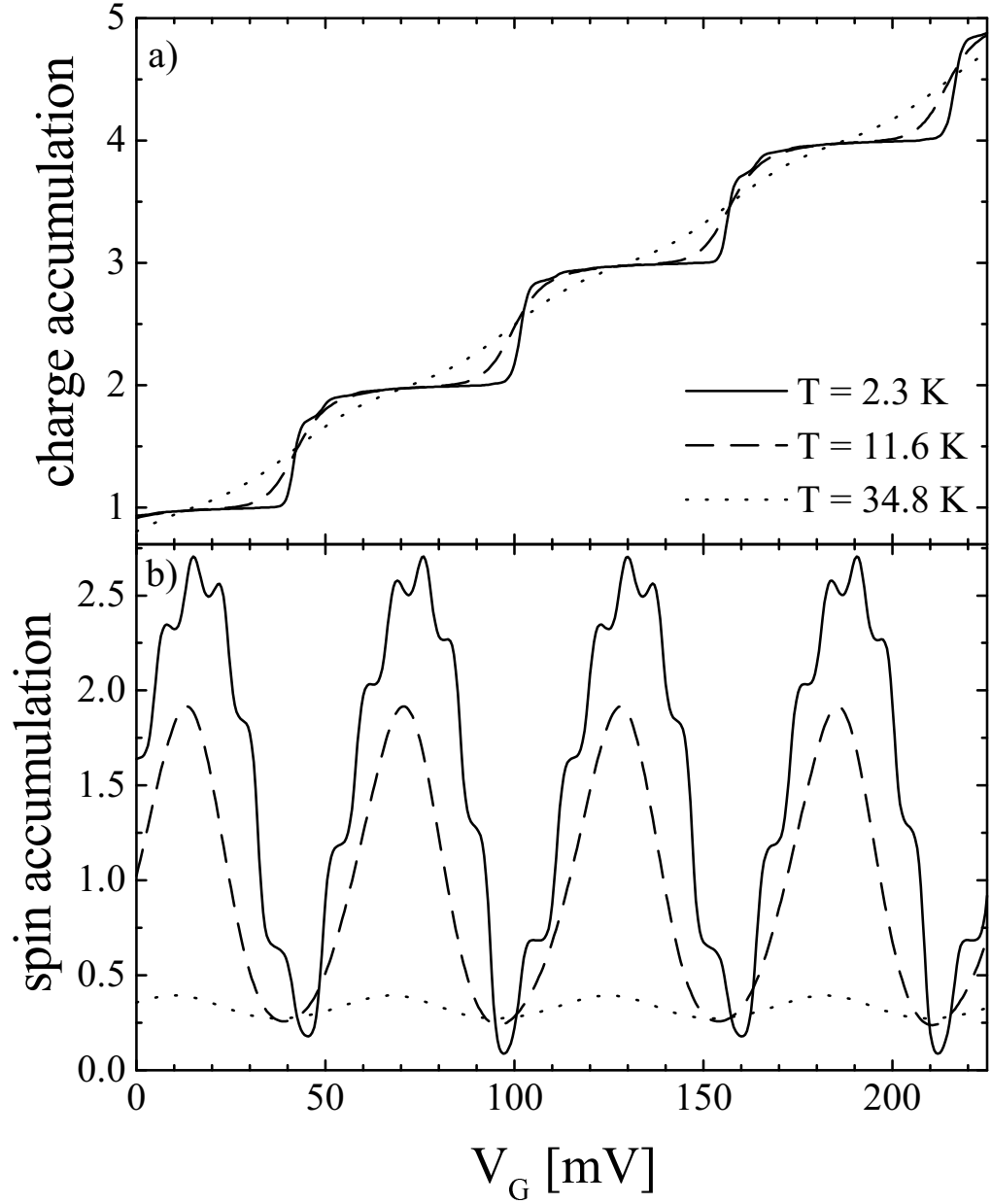 }

%\hskip 1cm
 \vspace{1cm}
 \caption{The charge (a) and spin accumulation (b) as a function
of the gate voltage $V_g$  for $V=15\;$mV and for $T=2.3$ K (solid
curve), $T=11.6$ K (dashed curve), $T=34.8$ K (dotted curve). The
other parameters of the system are as in Fig.2.} \label{Fig.10}
\end{figure}

\end{document}